\documentclass[prl,superscriptaddress,amsmath,amssymb,nofootinbib,twocolumn,floatfix]{revtex4}

\usepackage{setspace}
\usepackage{graphicx}
\usepackage{hyperref}

\usepackage{xcolor}

\newcommand{\sectionname}[1]{ \noindent{\bfseries #1}.---}

\begin{document}

\begin{flushleft}
KCL-PH-TH-2021-60
\end{flushleft}


\title{Search for a scalar induced stochastic gravitational wave background in the third LIGO-Virgo observing run}

\author{Alba Romero-Rodr\'\i guez}
\affiliation{Institut de F\'\i sica  d'Altes Energies (IFAE), Barcelona Institute of Science and Technology, E-08193 Barcelona, Spain}
\author{Mario Mart\'\i nez}
\affiliation{Institut de F\'\i sica  d'Altes Energies (IFAE), Barcelona Institute of Science and Technology, E-08193 Barcelona, Spain}
\affiliation{Catalan Institution for Research and Advanced Studies (ICREA), E-08010 Barcelona, Spain}
\author{Oriol Pujolàs}
\affiliation{Institut de F\'\i sica  d'Altes Energies (IFAE), Barcelona Institute of Science and Technology, E-08193 Barcelona, Spain}
\author{Mairi Sakellariadou}
\affiliation{Theoretical Particle Physics and Cosmology Group, \, Physics \, Department, \\ King's College London, \, University \, of London, \, Strand, \, London \, WC2R \, 2LS, \, UK}
\author{Ville Vaskonen}
\affiliation{Institut de F\'\i sica  d'Altes Energies (IFAE), Barcelona Institute of Science and Technology, E-08193 Barcelona, Spain}

\date{\today}


\begin{abstract}
Formation of primordial black holes from inflationary fluctuations is accompanied by a scalar induced gravitational wave background. We perform a Bayesian search of such background in the data from Advanced LIGO and Virgo's first, second and third observing runs, parametrizing the peak in the curvature power spectrum by a log-normal distribution. The search shows no evidence for such a background. We place 95\% confidence level upper limits on the integrated power of the curvature power spectrum peak which, for a narrow width, reaches down to $0.02$ at $10^{17}\,{\rm Mpc}^{-1}$. The resulting constraints are stronger than those arising from BBN or CMB observations. In addition, we find that LIGO and Virgo, at its design sensitivity, and the Einstein Telescope can compete with the constraints related to the abundance of the formed primordial black holes.
\end{abstract}

\maketitle


\sectionname{Introduction}
In addition to gravitational waves (GWs) from compact binary coalescences (CBCs)~\cite{LIGOScientific:2018mvr,LIGOScientific:2020ibl}, Advanced LIGO~\cite{LIGOScientific:2014pky} and Advanced Virgo~\cite{VIRGO:2014yos} can probe stochastic gravitational wave (GW) backgrounds providing invaluable information about the earliest stages of our universe. Such backgrounds originate for example from cosmic inflation, cosmological phase transitions or cosmic strings (see e.g.~\cite{Caprini:2018mtu,Christensen:2018iqi} for comprehensive reviews). Recently, the latest LIGO-Virgo results on searches for stochastic GW signals, using the first,  second and third (O1-O3) observation periods~\cite{LIGOScientific:2021yuo}, have been expressed in terms of constraints on cosmological phase transitions~\cite{Romero:2021kby} and cosmic strings~\cite{LIGOScientific:2021nrg}. In this Letter we focus on stochastic GW backgrounds accompanying formation of primordial black holes (PBHs) in the early universe, and present the results from a search of such signals in the LIGO-Virgo data~\cite{LVdata}.

PBHs have been subject to growing interest in recent years as the traditional particle dark matter candidates have become increasingly tightly constrained. Particularly interesting is the asteroid mass PBH mass window which currently remains unconstrained (see e.g.~\cite{Carr:2020gox} for recent review). Yet, even extremely light PBHs that have evaporated through Hawking radiation~\cite{Hawking:1975vcx} soon after their formation can have interesting consequences on formation of dark matter and baryon asymmetry in the universe~\cite{Allahverdi:2017sks,Lennon:2017tqq,Morrison:2018xla,Hooper:2019gtx,Baldes:2020nuv,Hooper:2020otu,Datta:2020bht}. Several studies have also been motivated by the LIGO-Virgo observations, investigating the possibility that at least some of the detected CBC events originate from PBH coalescences~\cite{Sasaki:2016jop,Bird:2016dcv,Clesse:2016vqa,Raidal:2017mfl,Ali-Haimoud:2017rtz,Raidal:2018bbj,Vaskonen:2019jpv,Gow:2019pok,DeLuca:2020qqa,Hall:2020daa,Wong:2020yig,Hutsi:2020sol,Boehm:2020jwd,Hutsi:2021nvs,DeLuca:2021wjr,Franciolini:2021tla}\footnote{Also the prospects of probing GW signals from PBH binaries with future detectors has been considered~\cite{Miller:2020kmv,Mukherjee:2021itf,Mukherjee:2021ags,DeLuca:2021hde,Pujolas:2021yaw}.}. In this Letter we will focus on another GW source related to PBHs, namely the large curvature fluctuations which lead to PBH formation. 

The most studied PBH formation mechanism relies on large curvature fluctuations generated during cosmological inflation~\cite{Hawking:1971ei,Carr:1974nx,Carr:1975qj}. As such, the PBH formation is accompanied with a strong stochastic GW background generated at the second order in the cosmological perturbation theory from scalar perturbations~\cite{Matarrese:1993zf,Matarrese:1997ay,Nakamura:2004rm,Ananda:2006af,Baumann:2007zm,Kohri:2018awv}. The formation of these scalar induced GWs has been extensively studied and their spectrum is well established -- in particular, Refs.~\cite{Inomata:2019yww,DeLuca:2019ufz,Yuan:2019fwv} recently showed that the leading order result for the scalar induced GWs formed during radiation dominated expansion is gauge independent. 

The spectrum of the scalar induced GWs is fixed by the curvature power spectrum. From the cosmic microwave background observations we know that at large scales the amplitude of the curvature power spectrum is $\mathcal{O}(10^{-9})$~\cite{Planck:2018vyg}. For such small amplitude curvature fluctuations the scalar induced GW background is extremely weak and can not be probed with any foreseeable experiment. However, for PBH formation the curvature power spectrum amplitude needs to be $\mathcal{O}(0.01)$ at some small scales (assuming PBH formation in the standard radiation dominated expansion). This corresponds to a sizeable scalar induced GW background within the reach of GW observatories~\cite{Saito:2008jc,Assadullahi:2009jc,Bugaev:2010bb,Alabidi:2012ex,Inomata:2016rbd,Orlofsky:2016vbd,Espinosa:2018eve,Inomata:2018epa,Byrnes:2018txb,Clesse:2018ogk,Wang:2019kaf,Chen:2019xse,Lewicki:2021xku}. Interestingly, the recent NANOGrav result~\cite{NANOGrav:2020bcs}, if interpreted as a stochastic GW background detection at nanoHerz frequencies, can be explained by scalar induced GWs from PBH formation~\cite{Vaskonen:2020lbd,DeLuca:2020agl,Kohri:2020qqd}. However, more data is needed in order to claim for a detection of a stochastic GW background.

Peaks in the curvature power spectrum that reach the amplitude $\mathcal{O}(0.01)$ required for PBH formation can be generated by features or turns in the inflaton potential. For example, an inflection point in single field inflation~\cite{Bullock:1996at,Yokoyama:1998pt,Garcia-Bellido:2017mdw,Ezquiaga:2017fvi,Kannike:2017bxn,Germani:2017bcs,Ballesteros:2017fsr,Hertzberg:2017dkh,Atal:2019erb,Ballesteros:2020qam}, the beginning of a waterfall phase in hybrid inflation~\cite{Garcia-Bellido:1996mdl,Lyth:2011kj,Clesse:2015wea,Inomata:2018cht}, or the change of the potential from convex to concave in thermal inflation~\cite{Dimopoulos:2019wew,Lewicki:2021xku} can trigger such growth of curvature fluctuations at small scales. 

In what follows we will not consider any particular model of inflation. Instead, we parametrize the peak in the curvature power spectrum by its position, width and integrated power. We perform a search of a scalar induced GW background in the LIGO-Virgo data generated by such peaks in the curvature power spectrum, and study whether LIGO-Virgo observations constrain the aforementioned PBH formation scenario. In Ref.~\cite{Kapadia:2020pnr} a similar analysis with data from LIGO second observing run has been presented.

\sectionname{Parametrization}
We choose to study a log-normal shape for the peak in the curvature power spectrum,
\begin{equation} \label{eq:LNpeak}
\mathcal{P}_\zeta(k) = \frac{A}{\sqrt{2\pi} \Delta} \exp\left[- \frac{\ln^2(k/k_*)}{2\Delta^2} \right] \,,
\end{equation}
where $A$ is the integrated power of the peak, $\Delta$ determines its width and $k_*$ its position.
In the $\Delta \to 0$ limit the spectrum reduces to a Dirac delta function, $\lim_{\Delta\to 0} \mathcal{P}_\zeta(k) = A \,\delta(\ln(k/k_*))$. At $\Delta\ll 1$ the amplitude of the induced GWs as well as the generated PBH abundance become independent of $\Delta$, whereas for $\Delta\gtrsim 1$ they are determined by the peak amplitude $\mathcal{P}_\zeta(k_*) = A/(\sqrt{2\pi} \Delta)$.

The expression for the scalar induced GW spectrum in terms of the curvature power spectrum is given e.g. in Refs.~\cite{Kohri:2018awv,Inomata:2019yww}. This spectrum is peaked around the same wavenumber as the curvature power spectrum, corresponding to the frequency $f_*/{\rm Hz} = 1.6\times 10^{-15} k_*/{\rm Mpc}^{-1}$, and its peak amplitude is $\Omega_{\rm GW} = \mathcal{O}(10^{-5}) A^2$ in the $\Delta\ll1$ limit. The LIGO-Virgo detectors, being sensitive to frequencies around $f\gtrsim 10\,{\rm Hz}$, have the potential to probe peaks in the curvature power spectrum at scales larger than $10^{15}\,{\rm Mpc}^{-1}$. Such scales re-entered the horizon at very high temperatures, $T \gtrsim 10^8\,{\rm GeV}$, so, assuming the Standard Model value for the effective number of relativistic degrees of freedom, we take $g_{*,s} = g_* \approx 100$.

\sectionname{Bayesian search}
The current estimates of the foreground from compact binary coalescence events (CBC) show that it constitutes a non-negligible component of any stochastic GW  signal~\cite{LIGOScientific:2017zlf,LIGOScientific:2021yuo}. We perform a simultaneous Bayesian search for a stochastic GW signal sourced by both scalar fluctuations and CBCs. The CBC component is well-approximated by a power law~\cite{Callister:2016ewt}, with a spectrum given by $\Omega_{\rm CBC} = \Omega_{\rm ref} (f/f_{\rm ref})^{2/3}$ where we take $f_{\rm ref}=25~\rm{Hz}$. The data is fitted to a model $\Omega_{\rm GW}(f, \boldsymbol{\theta})$ that includes four parameters, $\boldsymbol{\theta}=(\Omega_{\rm ref},A,k_*,\Delta)$.  Table~\ref{tab:priors} collects the priors used for each of the variables. The priors for $A$ and $k_*$ are chosen such that the resulting peak in the GW spectrum is compatible with the LIGO-Virgo sensitivity window both in amplitude and in frequency. The prior on $\Delta$ is instead chosen such that the range covers both very narrow and broad spectra. In the case of $\Omega_{\rm ref}$, the priors reflect the current estimates of the CBC background including uncertainties on the mass and redshift distributions of the CBC events~\cite{LIGOScientific:2017zlf,LIGOScientific:2021yuo}.
%

\begin{table}
\begin{center}
\begin{footnotesize}
\begin{tabular}{c| c} \hline
    {Parameter} & {Prior}\\
    \hline
    $\Omega_{\rm ref}$ & LogUniform($10^{-10}$, $10^{-7}$)\\
    $A$  & LogUniform($10^{-3}$, $10^{0.5}$)\\
    $k_*/$Mpc$^{-1}$ & LogUniform($10^{13}$, $10^{21}$)\\
    $\Delta$ & LogUniform($0.05$, $5$)\\
    \hline
  \end{tabular}
  \end{footnotesize}
  \caption{Prior distributions used for the amplitude of the CBC background at 25\,Hz, $\Omega_{\rm ref}$, the integrated power $A$ of the peak in the curvature power spectrum, its position $k_*$ and its width $\Delta$.}
  \label{tab:priors}
  \end{center}
\end{table}

 The Bayesian search follows that in Refs.~\cite{Mandic:2012pj,Callister:2017ocg,Meyers:2020qrb} with Gaussian log-likelihood for a single detector pair, 
%
\begin{equation} \label{eq:pe_ms:likelihood}
\ln p(\hat C_{IJ} | \boldsymbol{\theta}, \lambda) \propto \sum_{i} \frac{\left[\hat C_{IJ}(f_i) - \lambda \Omega_{\rm GW}(f_i, \boldsymbol{\theta}) \right]^2}{\sigma_{IJ}^2(f_i)} \,, 
\end{equation}
where the sum runs over the frequency bins $f_i$, $\hat C_{IJ}(f_i)$ is the cross-correlation estimator of the stochastic GW background calculated using data from detectors $I$ and $J$, $\sigma^2_{IJ}(f_i)$ is the corresponding variance~\cite{Allen:1997ad}, and 
$\lambda$ stands for the calibration uncertainties of the detectors \cite{Sun:2020wke}, which is marginalized over~\cite{Whelan:2012ur}. 
We analyse the O1-O3 LIGO-Virgo data, for which we perform a multi-baseline study summing the 
corresponding log-likelihoods for individual pairs of detectors.
Since the search for an isotropic stochastic signal shows no evidence of  correlated magnetic noise, and a pure Gaussian noise model is still preferred by the data~\cite{LIGOScientific:2021yuo},  a contribution from Schumann resonances~\cite{Thrane:2013npa,Coughlin:2018tjc,Meyers:2020qrb} is neglected. 

\begin{figure}
\begin{center}
\includegraphics[width=0.98\columnwidth]{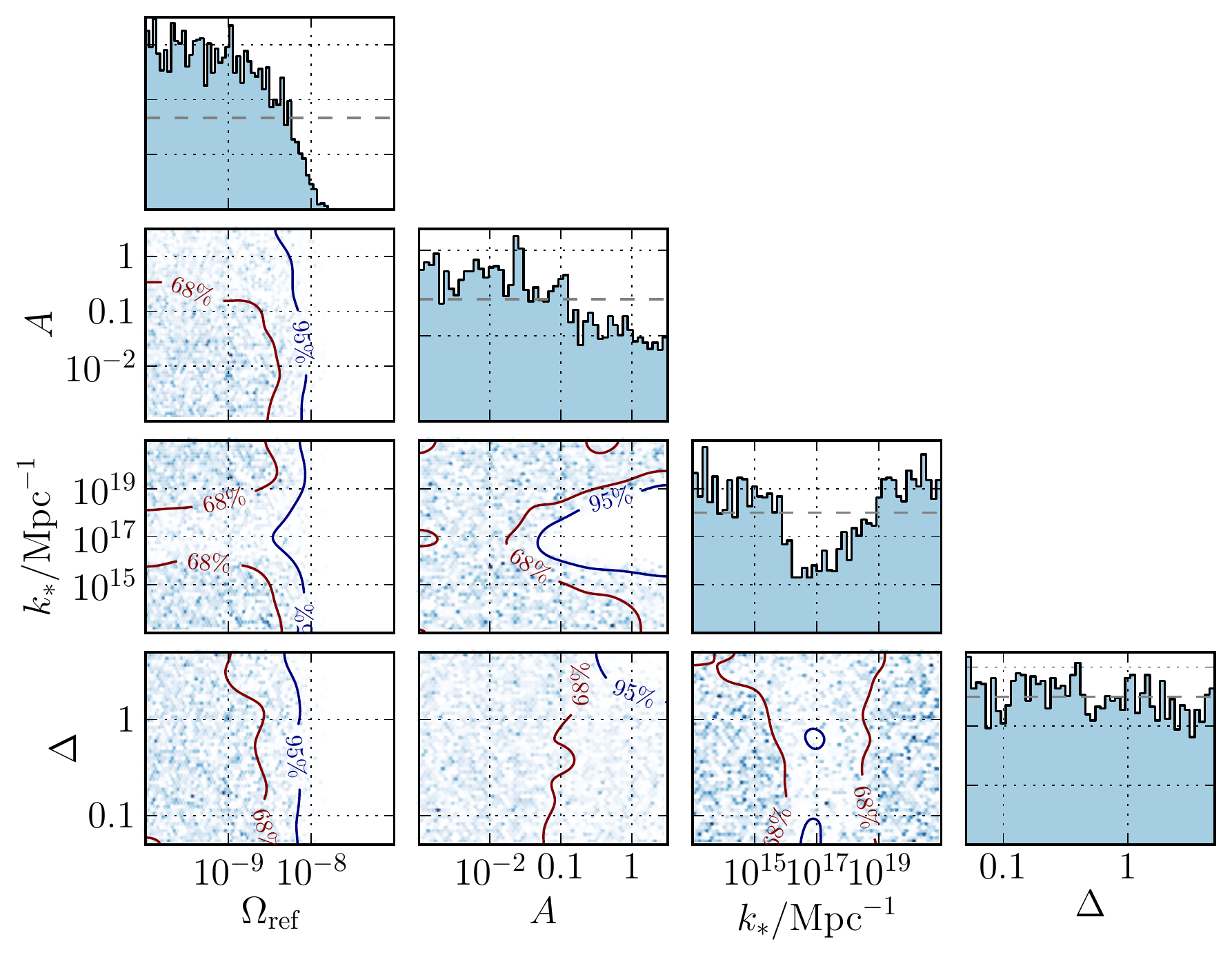}
\end{center}
\caption{Posterior distributions for the combined CBC and scalar induced GW search. The 68$\%$ and $95\%$ CL exclusion contours are shown. The horizontal dashed lines in the  histograms 
indicate the flat priors used in the analysis.}
\label{fig:corner_T}
\end{figure}

The results from the Bayesian search and the posterior distributions of the different parameters are presented in Fig.~\ref{fig:corner_T}. 
A Bayes factor of $\ln \mathcal{B}_{\rm noise}^{\rm CBC+scalar} = -0.8$ is obtained, indicating no evidence for a stochastic signal, as described by the combined CBC and scalar induced GW model,  in the LIGO-Virgo data.  A 95$\%$ confidence level (CL) upper limit on $\Omega_{\rm CBC}(25 \rm{Hz}) = 6.0\times 10^{-9} $ was obtained, compatible with the results obtained from other searches~\cite{Romero:2021kby}. The data excludes part of the parameter space in $k_*/{\rm Mpc}^{-1}$ and $A$, and this exclusion shows significant dependence on $\Delta$. The 95\% CL upper limits on  $A$ computed for given values of $\Delta$ and $k_*$ are collected in Table~\ref{tab:extra_runs}.
As expected, for large widths, $\Delta\gg 1$, the obtained upper limits on $A$ do not depend on the peak position, whereas for $\Delta\ll 1$ the most stringent bound on $A$ is obtained at the peak frequency near the best sensitivity of LIGO-Virgo detectors. The strongest exclusion, $A<0.02$ at 95\% CL, is obtained for a narrow peak at $k_* = 10^{17}\,{\rm Mpc}^{-1}$.

\begin{table}
\begin{center}
\begin{footnotesize}
\begin{tabular}{c|c|c|c} \hline
    & $k_*=10^{15}\, \rm{Mpc}^{-1}$ & $k_*=10^{17}\, \rm{Mpc}^{-1}$ & $k_*=10^{19}\, \rm{Mpc}^{-1}$ \\
    \hline
    $\Delta=0.05$  & $2.1$ & $0.02$ & $1.4$ \\
    $\Delta=0.2$  & $2.2$ & $0.03$ & $1.6$ \\
    $\Delta=1$  & $1.6$ & $0.05$ & $1.8$ \\
    $\Delta=5$ & $0.2$ &  $0.2$ & $0.3$ \\
    \hline
     \end{tabular}
  \end{footnotesize}
  \caption{Upper limits on the integrated power $A$ of the peak in the curvature power spectrum at 95\% CL for fixed values of the peak position $k_*$ and width $\Delta$.}
  \label{tab:extra_runs}
  \end{center}
\end{table}

\begin{figure}[h!]
\begin{center}
\includegraphics[width=0.98\columnwidth]{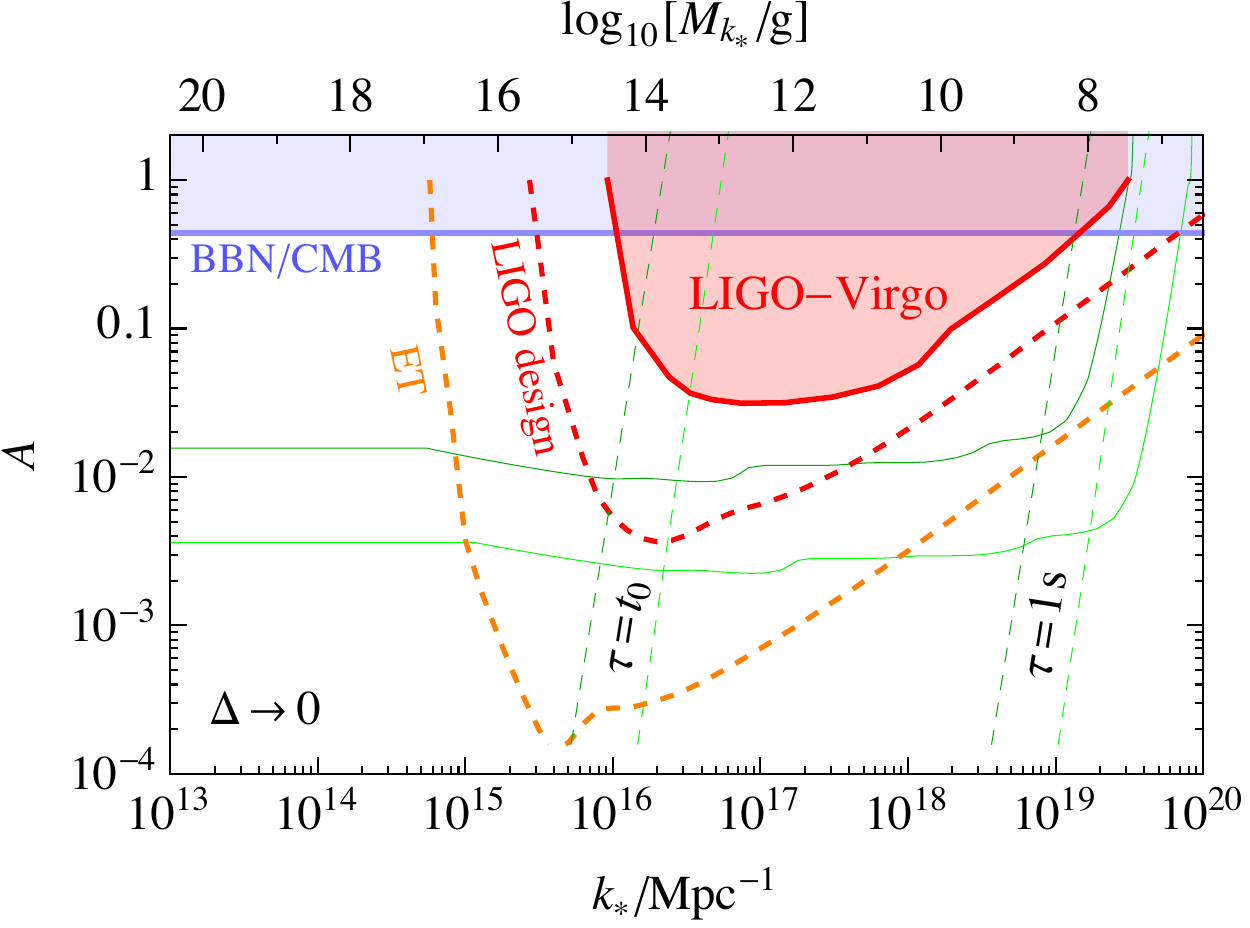} \\ \vspace{2mm}
\includegraphics[width=0.98\columnwidth]{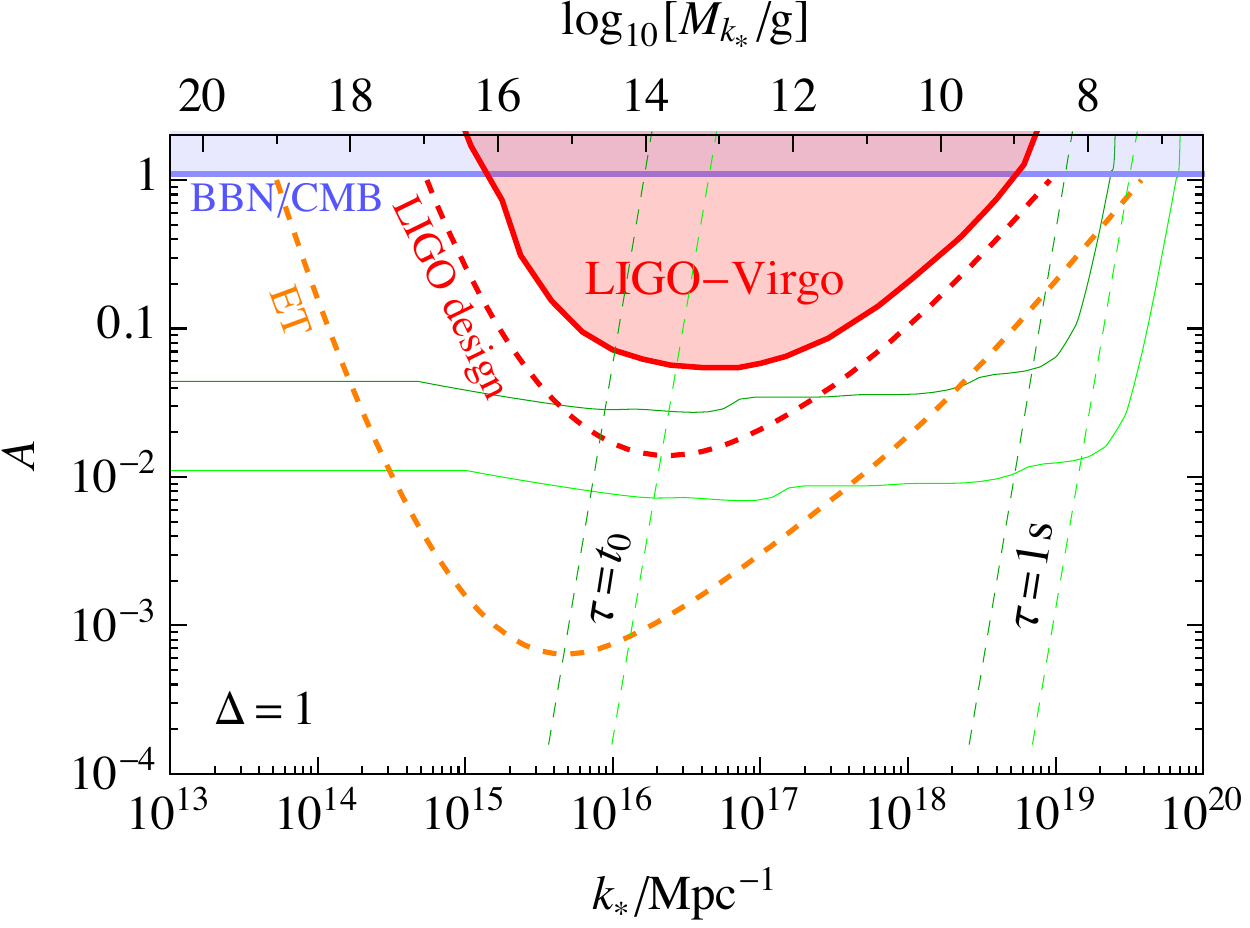}
\end{center}
\caption{Comparison between the LIGO-Virgo $95\%$ CL exclusions (red solid lines) for $\Delta \rightarrow 0$ (top) and $\Delta = 1$ (bottom) and the constraints arising from BBN/CMB (blue lines) and PBH formation (green lines, with the range between them indicating uncertainties in the calculation of the PBH abundance) on the integrated power of the curvature power spectrum (see text). The dashed red and orange curves show the projected sensitivities of the LIGO in its final phase and Einstein telescope (ET). The green dashed curves indicate the evaporation timescales of the PBHs.}
\label{fig:AkLIGO}
\end{figure}

\sectionname{Implications}
We discuss now the implications for PBHs, possibly formed by collapse of overdense Hubble patches. In Fig.~\ref{fig:AkLIGO} we show the current 95\% CL LIGO-Virgo bound for the peak integrated power $A$ of the curvature power spectrum as a function of the peak wavenumber $k_*$ obtained from the Bayesian analysis for a Dirac delta function peak ($\Delta \rightarrow 0$) and for a log-normal peak with $\Delta = 1$. In the range $\mathcal{O}(10^{15}) < k_*/{\rm Mpc}^{-1} < \mathcal{O}(10^{18})$ the LIGO-Virgo bound is stronger than the indirect bounds on the abundance of GWs arising from big bang nucleosynthesis (BBN), $\Omega_{\rm GW}h^2 < 1.8\times 10^{-6}$~\cite{Kohri:2018awv}, and the cosmic microwave background (CMB) observations, $\Omega_{\rm GW}h^2 < 1.7\times 10^{-6}$~\cite{Pagano:2015hma}.

To compare the LIGO-Virgo bound on $A$ to the constraints induced by PBH formation, we calculate the PBH abundance generated from the peak in the curvature power spectrum of the form of Eq.~\eqref{eq:LNpeak}. We follow the calculation presented in Ref.~\cite{Young:2019yug} of the PBH formation taking into account the critical collapse~\cite{Choptuik:1992jv,Niemeyer:1997mt,Niemeyer:1999ak}, and the non-linear relation between the density contrast and the curvature perturbation, $\delta_{m} = \delta_{\zeta} - 3\delta_{\zeta}^2/8$~\cite{Young:2019yug,DeLuca:2019qsy,Kawasaki:2019mbl}. We assume that the curvature perturbations are Gaussian and relate their variance to the curvature power spectrum using a real-space top-hat window function accounting for the damping of sub-horizon fluctuations with linear transfer function~\cite{Young:2019yug}. The masses of the PBHs follow the scaling law $M = \kappa M_k (\delta_m - \delta_c)^\gamma$, where $M_k$ is the horizon mass at the horizon re-entry time of scale $k$, $\gamma=0.36$ is the universal critical exponent~\cite{Choptuik:1992jv,Evans:1994pj} and $\delta_c$ is the collapse threshold parameter. The precise value for the threshold is uncertain, but at present it is believed to be in the range $\delta_c \in (0.41,2/3)$~\cite{Harada:2013epa,Musco:2018rwt}. We will consider the following nearly extremal values for $\delta_c$ and $\kappa$~\cite{Young:2019yug}: i) $\kappa = 11.0$ and $\delta_c = 0.45$, ii) $\kappa = 3.0$ and $\delta_c = 0.65$. The differences in the resulting PBH mass distribution and abundance in these two cases reflect the uncertainties in the calculation of PBH formation~\cite{Young:2019yug,Young:2019osy,Young:2020xmk,Gow:2020bzo,Musco:2018rwt,Escriva:2020tak}.

PBHs produced by critical collapse form when large enough curvature perturbations of scale $k$ re-enter the horizon. This leads to the relation between the mean PBH mass and the wavenumber, $M_{\rm PBH} \sim M_k$, 
which is used in Fig.~\ref{fig:AkLIGO} to display the PBH mass associated with modes of different $k$ in the upper horizontal axis~\footnote{Notice that this is only a rough estimate of the mean PBH mass. In fact, following the critical collapse scaling law, it depends weakly on the amplitude of the fluctuations.}. For the sensitivity range of LIGO-Virgo detectors, the relevant PBH masses are $M_{\rm PBH} \lesssim 10^{16}$\,g. Such PBHs evaporate very efficiently through Hawking radiation~\cite{Hawking:1974rv,Hawking:1975vcx}. The abundance of PBHs that currently are strongly radiating is strictly constrained by extragalactic $\gamma$-ray background observations to be much smaller than the dark matter abundance~\cite{Carr:2009jm}. The abundance of lighter PBHs that have evaporated completely by the present time, $M_{\rm PBH} \lesssim 10^{15}$\,g, right from the green dashed lines labelled as $\tau = t_0$, is constrained by CMB observations and by their effects on BBN~\cite{Carr:2009jm}. The abundance of PBHs of mass $M_{\rm PBH} \lesssim 10^{9}$\,g, right from the green dashed lines labelled as $\tau = 1\,$s, is unconstrained, as they would have completely evaporated before BBN.

In Fig.~\ref{fig:AkLIGO} we show the envelope of the collection of these constraints by the green curves~\footnote{The constrains of \cite{Carr:2009jm} are converted for our non-monochromatic PBH mass functions using the method of Ref.~\cite{Carr:2017jsz}.}.  As discussed above, we consider two cases which reflect the theoretical uncertainties in the PBH formation. The upper green curve corresponds to $\kappa = 3.0$ and $\delta_c = 0.65$, and the lower one to $\kappa = 11.0$ and $\delta_c = 0.45$. We see that the current 95\% CL LIGO-Virgo bound lies above both of these curves, and therefore we conclude that currently still the LIGO-Virgo sensitivity is not enough to constrain the PBH formation~\footnote{The PBH abundance calculation in Ref.~\cite{Kapadia:2020pnr} was done using a Gaussian instead of a real-space top-hat window function with a different value of the critical threshold. We emphasize that our calculation consistently follows Ref.~\cite{Young:2019yug}. We believe that this causes the difference in their results as compared to ours.}. The dashed contours instead show the projected exclusion for Advanced LIGO at its designed sensitivity~\cite{LIGOScientific:2014pky} and that of te next generation experiment Einstein telescope (ET)~\cite{Punturo:2010zz}. These are calculated simply by requiring a signal-to-noise ratio for signal detection above eight~\footnote{We note that this may, in particular for narrow GW spectra, slightly overestimate the sensitivity.},  and indicate that in near future the ground-based GW detectors coul probe PBH formation.

\sectionname{Conclusions}
We have searched for a stochastic GW background created by large scalar fluctuations in the early universe,  using LIGO-Virgo O1, O2 and O3 public cross-spectrum data. Without assuming any particular model for inflation, we have described the curvature power spectrum using a log-normal function describing the main features in terms of integrated power, position and width of a peak in the curvature power spectrum. A Bayesian analysis, including also potential contributions from CBC astrophysics foregrounds, did not show evidence for such a GW background, but indicated that the data has the sensitivity to exclude part of the parameter space of the model. In particular, we have derived constraints, depending on the width of the peak, on the integrated power of the peak as a function of its position. We have shown that these constraints are stronger than the ones arising form BBN and CMB observations in the range $\mathcal{O}(10^{15}) < k_*/{\rm Mpc}^{-1} < \mathcal{O}(10^{18})$. These constraints, reaching $A\simeq 0.02$ for a narrow peak at $k_*\simeq 10^{17}\,{\rm Mpc}^{-1}$, are still not strong enough to compete with the constraints arising from the abundance of PBHs that such peak in the curvature power spectrum corresponds to. However, we find that current ground-based experiments at their design performance, and the future Einstein Telescope will reach the required sensitivity, providing a very powerful probe of the standard formation mechanism of very light PBHs.

\sectionname{Acknowledgements}
The authors would like to thank T. Callister for his assistance with the Bayesian search code. 
The authors are grateful for computational resources provided by the LIGO Laboratory and supported by National Science Foundation Grants PHY-0757058 and PHY-0823459. This paper has been given LIGO DCC number LIGO-P2100271 and Virgo TDS number VIR-0886A-21.  This work is partially  supported   by  the Spanish MINECO  under the grants SEV-2016-0588, PGC2018-101858-B-I00 and FPA2017-88915-P, some of which include ERDF  funds  from  the  European  Union, as well as grant 2017-SGR-1069 from Generalitat de Catalunya. IFAE is  partially funded by the CERCA program of the Generalitat de Catalunya. M.S. is supported in part by the Science and Technology Facility Council (STFC), United Kingdom, under the research grant ST/P000258/1.


\bibliography{pbh}

\end{document}